\documentclass[journal]{IEEEtran}
\usepackage{color}
\usepackage{url}

\usepackage{algorithm}
\usepackage{algorithmic}
\usepackage{ragged2e}
\usepackage[pdftex]{graphicx}
\DeclareGraphicsExtensions{.pdf, .pdf.bb, .eps}

\usepackage{amssymb}
\usepackage{cite}

\ifCLASSOPTIONcompsoc
  \usepackage[nocompress]{cite}
\else
  \usepackage{cite}
\fi

\usepackage{amsmath}

\ifCLASSINFOpdf
\else
\fi
\begin{document}

%
\title{Analysis of   Bitcoin Vulnerability to  Bribery Attacks Launched Through Large Transactions}
%
%
%

\author{Ghader~Ebrahimpour,~\IEEEmembership{Member,~IEEE,}
        Mohammad~Sayad Haghighi,~\IEEEmembership{Senior Member,~IEEE}
\thanks{Manuscript was initially developed in March 2020.}
\thanks{Ghader Ebrahimpour is with the School
of Electrical and Computer Engineering, University of Tehran,Iran
 e-mail: g.ebrahimpour@ut.ac.ir.}
\thanks{Mohammad Sayad Haghighi (corresponding author) is with the School of  Electrical and Computer Engineering, University of Tehran, Iran, e-mail:  sayad@ut.ac.ir, sayad@ieee.org.}
}

%
%

\markboth{Draft v14, Jan~2021}%
{Shell \MakeLowercase{\textit{et al.}}: Bare Demo of IEEEtran.cls for IEEE Journals}
%



\maketitle

\begin{abstract}
Bitcoin uses blockchain technology to maintain  transactions order and provides probabilistic guarantee to prevent double-spending, assuming that an attacker's computational power does not exceed \%50 of the network power. In this paper, we design a novel bribery attack and  show that this guarantee can be hugely undermined. Miners are assumed  to be rational in this setup and they are given incentives that are dynamically calculated. In this attack, the  adversary misuses the Bitcoin protocol to  bribe miners and  maximize their gained advantage.  We will reformulate  the bribery attack  to propose a general mathematical foundation upon which we build multiple strategies. We show that,  unlike   Whale Attack,  these strategies are practical. If the rationality assumption holds, this shows how vulnerable blockchain-based systems like Bitcoin  are. We suggest a soft fork on Bitcoin to fix this issue at the end.
\end{abstract}


%
\IEEEpeerreviewmaketitle

\section{Introduction}\label{sec:introduction}
%
%
%
%
\IEEEPARstart{B}{itcoin}  was introduced in 2008 as a peer to peer electronic currency and its aim was to make online transactions outside the classic financial system possible~\cite{nakamoto2019bitcoin}. This was one of the first practical efforts toward bringing trust to transactions in a zero-trust environment \cite{arabsorkhi2016conceptual,jafarian2020discrimination}. Study of the digital currency history before that \cite{narayanan2015bitcoin} shows  decades of research did not lead to a large-scale deployment of e-cash, and with the emergence of   Bitcoin, all those efforts  have been dwindled~\cite{chaum1983blind, okamoto1989disposable, okamoto1991universal, trolin2005universally, chaum1988untraceable, camenisch2005compact}.  Blockchain \cite{haber1990time}, as the core of Bitcoin, is a data structure to save the history of transactions in a distributed system. Each block in blockchain uses another data structure called Merkle tree~\cite{merkle1979secrecy} to store transactions. 

Bitcoin provides probabilistic guarantee to prevent double-spending, assuming that no attacker's computational power  exceeds \%50 of the network mining power. In this paper, we design a novel attack scenario to show that Bitcoin's probabilistic guarantee on double spending can be hugely undermined. Miners are expected to be rational and work based on some monetary incentives. In our  proposal, the attacker misuses Bitcoin standard mechanisms to bribe   miners and  increase their advantage in a way that they  are encouraged  to follow the attacker's lead. We will show that this new attack can be  more efficient, practical and cheaper than the Whale Attack if the attacker adopts right strategies in distributing the bribe money even when paying the bribe out of the already spent money. We design a few of such strategies and analyze them analytically by using a Markov model of ours. The results show that this new attack can be quite challenging for block-based cryptocurrencies like Bitcoin and necessitate  taking  additional preventive  measures in the design of such cryptocurrencies. Two solutions  will also be proposed at the end of the paper that make soft-forks on Bitcoin.
In summary, the contributions of this work are:
\begin{itemize}
\item Proposing a new mathematical model for  Bribery attack and laying the foundations for further research.
\item Design  and evaluation of new strategies that can decrease the attack cost and increase the success probability.
\item Introduction of a commitment mechanism to make the attack more practical and decrease the attack cost and increase the success probability.
\end{itemize}

The rest of the paper is organized as follows. In Section~\ref{sec:related} we will review the state of the art attacks on Bitcoin.  In Section~\ref{sec:proposal} and~\ref{sec:analysis}, the proposed attack and its mathematical analysis will be presented. Numerical analysis of different attack strategies will be discussed in Section~\ref{sec:results}. The paper is finally concluded in Section~\ref{sec:conclusion}.

 

\section{Related Work\label{sec:related}}

Previous literature presented different types of attacks on Bitcoin.
Double spending  is the first and  the most common attack introduced in the field of cryptocurrencies \cite{karame2012double}, \cite{rosenfeld2014analysis}. In \%51 attack adversaries have control over the majority of the network hash power. A method to prevent such attacks is explored in \cite{bastiaan2015preventing}. Some types of attacks pertain to client-side applications, the security measures  they have adopted, users' mistakes, etc. Many Bitcoin users lost their money due to security breaches and poor key management of wallets \cite{krombholz2016other,haghighi2020computationally, haghighi2020intelligent}. 
In Sybil attack a single attacker  controls multiple nodes in the network or creates multiple identities by spawning up multiple virtual machines and IP addresses \cite{douceur2002sybil,haghighi2010neighbor,haghighi2019highly}. In \cite{zhang2019double}, a new attack for Bitcoin was introduced by combining double-spending  with  Sybil attack. In eclipse attack \cite{heilman2015eclipse}, an adversary takes control of a sufficient number of IP addresses and practically separates the victim (miner) from the public network. Stubborn Mining \cite{nayak2016stubborn} generalized Selfish-Mining \cite{eyal2014majority} by defining new strategies and combining them with eclipse attack to increase the  earnings of  miners. Block withholding attack was initially proposed by Rosenfeld in 2011. 
There are two types of block withholding attack \cite{rosenfeld2011analysis}. One, known as the Finney Attack, aims for financial gain when a double spending occurs. The second type is meant to cause financial harm to a pool operator. There are other attacks for  Bitcoin like Time Jacking attack~\cite{vyas2014security}, DDoS \cite{johnson2014game, haghighi2020intelligentss}, etc. which are categorized in four groups in [23]: Double spending, Wallet attacks, Network attacks, Mining attacks.

Bribery attack, which is  categorized in the same family as the attack studied in this paper, was introduced in~\cite{bonneau2016buy} and implemented in \cite{mccorry2018smart} as smart contracts to let a briber  exchange bribes with miners. In this attack, an adversary obtains the majority of computational power through bribing for a limited duration. This research article did not provide any modeling or analysis of the attack whatsoever. A way of giving incentives for adversarial purposes was proposed in \cite{liao2017incentivizing} under the name of Whale Attack. In  Whale Attack the attacker tries to issue whale transactions with high transaction fees to encourage miners to work on the forked chain. As mentioned in \cite{liao2017incentivizing}, the miners' revenue from the transaction fees does not exceed 0.00016BTC for each 257-byte transaction, unless the attacker uses exorbitant transaction fees\footnote{cc455ae816e6cdafdb58d54e35d4f46d860047458eacf1c7405dc6346-31c570d} to pay the bribery. A new study \cite{judmayer2019pay} uses out-of-band bribing to decrease the cost of attack and make it practical and efficient. However, its drawback is that the attacker loses  money if the attack fails. In in-band bribing, the attacker bribes out of the money  he/she has spent as in double-spending and thus, if the attack fails, the attacker loses nothing, except perhaps the computational power used during the attack. A more recent study \cite{sun2020model} tried to model  bribery attack by considering three parties; the attacker, the honest miners, the bribed miners. This study assumed that the bribed miners participate in a pool, and analyzed the bribed miners profit based on the computational power of the pool. It was shown that such an assumption would reduce the attack cost but creating such a pool can be impractical. Unlike this study, we analyze the attack based on the current pools in the network.

In the next section, we will deeply analyze the Bribery attack and show how effective this attack can be made by using the outputs of the proposed analysis framework.

\section{The Proposed Bribery Attack\label{sec:proposal}}
\subsection{Assumptions}
We assume that miners are rational. The rationality assumption is the expectation that miners will select, from a series of choices, the one(s) that will maximize their profit. We also assume that the distribution of mining power remains constant during the attack, and it is known to all parties. We normalize the network computational power  to 1 (or \%100). 
\subsection{Attack Process in a Nutshell}
Suppose that Alice, who controls a minority of the computational power in the Bitcoin network, attempts to launch a double-spending attack against a seller whose name is Bob. First, Alice does a large transaction with Bob. After the transaction is included in the chain, Alice starts working on a fork by creating a block that does not include that transaction, but keeps this matter private. Alice needs to create at least one block in her chain before the transaction is confirmed. This block has some special transactions in which Alice transfers some BTC to new addresses she has created before. After Alice's transaction to Bob is confirmed (generally, in 6 blocks time), Bob sends Alice the purchased goods, and Alice releases her block(s) and tries to "bribe" other miners to mine on her fork. To do this, Alice discloses the private key(s) (of one or more) of the account(s) that she has transferred the money to in the block(s) she has just created. Rational miners who see the private key(s), may decide to work on Alice's fork. They will create a new transaction to send the BTCs in the disclosed account to their accounts, and then try to create new blocks on Alice's block. As soon as a miner finds a solution, Alice confirms it by disclosing the next private key. If she manages to give sufficient incentives to attract enough mining power to her fork, the second branch will take over the main one and the transaction between Alice and Bob will be undone.

Unless Alice encourages other miners to work on her fork, she cannot hope for her attack to be successful, because she has a small fraction of the network computational power. If mining on Alice's fork is deemed more profitable by rational miners, they will join Alice's fork. She needs to keep convincing miners to mine on her fork, while making sure that the attack remains profitable for her at the end.

In order to initiate the attack, Alice needs at least one block in her fork that contains transactions to the account IDs to be published. Assuming that a transaction needs six blocks to be confirmed, the attacker should have at least \%14.28 of the network computational power. With such a power, the attacker can (in average) create one block in 70 minutes, that is the time required for the creation of 6 blocks by other miners (with \%85.72 of the power) in the main fork.

\section{Mathematical Analysis of the Attack\label{sec:analysis}}
We first describe a basic analysis of our attack by using a Markov model. Some parts of the analysis are similar to the work done in \cite{liao2017incentivizing} to analyze the Whale Attack. However, we diverge at some point as that study only includes a special case of the attack, compared to  our analysis which is generic and comprehensive. We  also introduce some practical strategies that Alice can adopt to reduce the attack cost and increase her success rate.

\subsection{A Basic Analysis}

In our analysis we will use the notations  listed in Table. \ref{table:syms}. We have:

\begin{table}
\caption{Symbols \& Notations}
\label{table:syms}
\vspace{-1ex}\begin{tabular}{|c|p{2.7in}|}
\hline
\bfseries Symbol & \bfseries Description\\
\hline\hline
$X, Y$ & The forked chain and the main chain, respectively.\\
\hline
$L(X,t)$ & The length of X at time t from the starting point of fork\\
\hline
$L(Y,t)$ & The length of $Y$ at time t from the starting point of fork\\
\hline
$\mu_i,\lambda_i$ & The mining power on $X$ and $Y$ at state $i,$ respectively\\
\hline
$P_i$ & The mining power of miner $i$\\
\hline
$Z$ & Set of all miners in descending order\\
\hline
$S$ & Set of all states in attack Markov chain\\

\hline
$R_t$ & {The amount of reward proposed by attacker at time t}\\
\hline
$F$ & {The Bitcoin mining reward (6.25 BTC at the time of writing)}\\
\hline
$C$ & {The number of blocks that are required for confirmation of a transaction (6 in practice)}\\
\hline
$l$ & {The number of blocks that the attacker mined in her chain before the the completion of transaction confirmation}\\
\hline
\end{tabular}
\end{table}

\begin{align}
\label{equ1}
\lambda = &\sum\limits_{\forall{i}}{P_i},\qquad \mu + \sum\limits_{\forall{i}}{\lambda_i} = 1
\end{align}

In the basic analysis, we denote Alice's forked chain by $X$ and the main chain by $Y$. We define  $D_t$ to be the difference between these chains:
\begin{equation}
\label{equ3}
D_t = L(Y, t) - L(X, t), \quad D_0 = C-l+1
\end{equation}

\begin{figure}[t]
        \centering
        \includegraphics[width=\columnwidth]{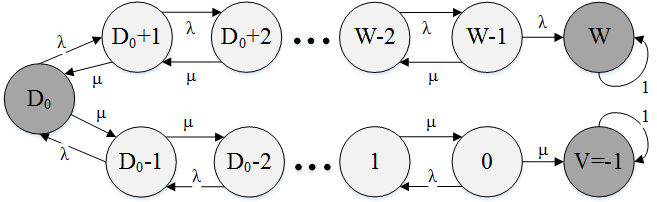}
        \vspace{-4ex}\caption{Markov chain model of a bribery attack in Bitcoin.}
        \label{fig:basemarkov}
\end{figure}

Based on \eqref{equ3}, we develop a model for Bitcoin as shown in Fig. \ref{fig:basemarkov}. The figure shows an absorbing Markov chain. State $V$ is the state in which all  miners are disappointed to mine on the main chain. State $W$ is similar to $V,$ but in that, all miners are mining on the main chain. Henceforth, we refer to finding a new block as 'event'.

Theorem 11.3 of \cite{grinstead2012introduction} states that in an absorbing Markov chain, the probability that the process is absorbed is equal to~1.
Based on this theorem, the attack eventually reaches either $W$ or $V$, that is, the attack either succeeds or fails. But there are some other interesting questions, including:

\begin{enumerate}
        \item On  average, how many times will the attack be in each transient state (states other than $W$ and $V$)?
        \item What is the probability that the attack  ends up in $V$, or alternatively, in $W$?
\end{enumerate}
\begin{equation}
\label{eqn_example}
Q_{BaseMarkov} = \begin{bmatrix}
Q & G \\
0& I \\
\end{bmatrix}\nonumber
\end{equation}

In our model, we have two absorbing states and $h$ transient states. The transition matrix of the chain is denoted by $Q_{BS}$ (or $Q_{BaseMatrix}$). We call this matrix the $BaseMarkov$. This matrix shows a Canonical Form \cite{grinstead2012introduction} of the probability matrix, {in which $I$ is a 2-by-2 identity matrix}, $0$ is a 2-by-$h$ zero matrix, $Q$ is an $h$-by-$h$ matrix and $G$ is a nonzero $h$-by-2 matrix. In $Q_{BS}$, the first $h$ states are transient and the last two ones are absorbing. Now, we can use Markov properties to answer the above questions. For the first one we use Theorem 11.4 of \cite{grinstead2012introduction} which states that in an absorbing Markov chain, the fundamental matrix for the chain is computed as:

\begin{equation}
\label{equ5}
N = (I-Q)^{-1}
\end{equation}
The entry $N_{ij}$ gives the expected number of times that the process is in the transient state $j$ if it is started from the transient state $i$. Based on this theorem it is sufficient to compute matrix $N$, and select the row $D_0$ (the initial state of the attack). The entries in this row will show the number of times that the attack passes each state before it succeeds or fails. Obtaining $N$ is very important for Alice, because she will use it as a parameter to design the attack process and determine the amount of reward/bribe that should be allocated to each step of the attack, so that it remains profitable for both the attacker and miners. Having $N,$ it is possible to obtain $e$ by using \eqref{equ6}.
\begin{equation}
\label{equ6}
e = Nc
\end{equation}
In the above, $c$ is a column vector whose entries are all $1$, and $e$ is a column vector whose $i$th entry is $e_i$.  $e_i$ is the expected number of steps taken before absorption assuming that {the chain starts from state $i$}.  This implies that  entry $D_0$ (of $e$) will be the average number of states the attack goes through before it  succeeds or fails.

To answer the second research  question on the probability of attack success and failure, we take advantage of Theorem {11.6} in \cite{grinstead2012introduction}. Assume $B$ is an $h$-by-$r$ matrix that is computed as:
\begin{equation}
\label{equ7}
B = NG
\end{equation}
where $G$ is  in the Canonical Form. Then, the entry $B_{ij}$ in $B$ is the probability that an absorbing Markov chain ends up in the absorbing state $j$, after starting from $i$. Using the result of this theorem, we can obtain the probability that the attack succeeds (V) or fails (W).

The cost of attack and its success can be obtained using these theorems. An attacker can design strategies and evaluate them using these theorems. Miners will also use the theorems to determine the profitable branch to mine on. To calculate the more profitable branch, we need to set some initial values:
\begin{eqnarray}
\lambda_i+\mu_i=1, \quad \lambda_{D_0}=\lambda, \quad \mu_{D_0}=\mu
\end{eqnarray}

In state $i ,$ Alice proposes $R_i$ as the reward of mining on $X$. At this time, a rational miner named $m$ with a mining power of $P_m$ wants to select the more profitable chain to mine on. If $m$ chooses $X$, then he receives the block reward only if the attack ultimately succeeds. Conditioned on the attack success,  $m$ receives the rewards with a probability of $P_m /(\mu_i + P_m)$. If $P_{X,S,i}$ is the probability  that the attack is successful at~$i$, and $P_{Y,F,i}$ is the probability of the attack failure at $i$, miner $m$, between $X$ and $Y, $ chooses the one that maximizes its expected profit. Mining on  chain $X$ is profitable for miner $m$ in state $i$ when:

\begin{equation}
\label{equ15}
R_i > \frac{P_{Y,F,i} \times (P_m + \mu_i)}{\lambda_i \times P_{X,S,i}} \times F-F
\end{equation}

To calculate $P_{Y,F,i}$ and $P_{X,S,i}$ in \eqref{equ15}, one can use Theorem 11.6 of~\cite{grinstead2012introduction}. For this purpose, we need the transition probability matrix which in turn needs the bribe distribution (i.e. the interdependence between probability matrix and bribe distribution). To overcome this interdependence, we need a strategy in distributing the bribe which is designed and explained in the next section.  In a special case, if the transition probability in the Markov chain remains constant during the attack time (worst case: none of the miners accepts the bribe), then  $m$ will use \eqref{equ16} to decide if $X$ is more profitable than $Y$. This special case is the whole thing that  the authors in \cite{liao2017incentivizing} wanted to reach.

\begin{equation}
\label{equ16}
R_i > \frac{(1-(\frac{\mu}{\lambda})^{i+1}) \times (P_m + \mu)}{\lambda \times (\frac{\mu+P_m}{\lambda-P_m})^{i+1}} \times F-F
\end{equation}

To attract and keep $m$ on  chain $X$, (\ref{equ15}) should be valid for all the states ($V<i<W$). To keep $m$ on  $X$, the necessary condition is:
\begin{equation}
\label{equ17}
\sum_{i=0}^{W-1}\frac{P_{X,S,i} \times P_m}{\mu_i+P_m} \times(R_i+F) > \sum_{i=0}^{W-1}\frac{P_{Y,F,i} \times P_m}{\lambda_i} \times F
\end{equation}
in which, unlike  \cite{liao2017incentivizing}, $\mu_i$ is not necessarily constant for all $i$ values. The sufficient condition is  to distribute $R_i$ in  a way that persuades $m$ to mine on $X$. A rudimentary way of attracting and keeping $m$ is to find $R_i$ for each state based on \eqref{equ15}. Nevertheless, Alice can distribute the bribes in different ways which result in different success probabilities and attack costs. We call the method of distributing the bribes, the attack strategy, which is studied next. We refer to \eqref{equ16} as the \textit{b\emph{\textit{asi}}c formula}. For example, if the attacker and the miner $m$'s computational powers are 20\% and 10\%\ (of all of the network power),  and $D_0 = 6$, then Alice should offer 876.2 BTC as the bribe before any event occurs. If the first event occurs on Alice's chain, then she should offer an additional bribe of 371.9 BTC  to keep her chain profitable to  miner $m$. The more events occur on her chain, the less bribe she needs to offer. Equation \eqref{equ18} shows the average attack cost regardless of its success or failure. Since in our scenario the attacker uses inline bribing, she will not lose anything in case of attack failure, thus, \eqref{equ19} shows the average cost of attack when it is successful ($AS$).

\begin{eqnarray}
\label{equ18}
&&AttackCost = \sum_{i=0}^{W-1}N_{D_0i} \times R_i\\
\label{equ19}
&&AttackCost_{AS} = \sum_{i=0}^{W-1}\frac{B_{iV}}{B_{D_0V}} \times N_{D_0i}\times R_i
\end{eqnarray}

In the equations, $N_{D_0i}$ is obtained from \eqref{equ6}, and $B_{ij}$ is obtained from \eqref{equ7}.
If Alice's goal is only to attract $m$ to her chain and if she visits each state at most once  (that is when all the events happen on chain $X$), the value of $AttackCost$ for the above example will become 1495.6 BTC. It is worth knowing that this attack will succeed with a probability of \%0.26. As Alice mines on her chain too, she will try to obtain the bribe money she has offered as an ordinary miner. This has not been noted in any of the previous studies as far as  we know.  So in our example, the expected reward that Alice will earn is 997.1 BTC and $m$'s reward will be 498.5 BTC.

As we saw above, in the \textit{basic analysis}, the cost of attack is very high and the probability of success is too low. These render the attack impractical. The reason  is that the analysis assumed that there was only one rational miner in the network and ignored others' computational power. To extend the analysis and to make the attack more practical, we need a bribing strategy.  In the following section we will propose some strategies that reduce the $AttackCost$ and increase the success probability.

\subsection{Attack Strategies}

To extend the analysis to a network in which more than one rational miner exist, we need bribing strategies. Here, we propose and analyze three:

\begin{itemize}
        \item 'Biggest Fish First while Keeping the Previous Catch (BFF)'     
                \item 'Constant-Rate Bribing' strategy (CRB).
        \item 'Guaranteed Variable-Rate Bribing with Commitment' strategy (GVC).
\end{itemize}

\subsubsection{BFF  Strategy}
In this strategy, and in each state of the Markov chain, Alice tries to attract the next biggest miner from  chain $Y$ and bring it to $X,$ while preserving the miners she has attracted in previous states. In the following theorem, we will prove that selecting the biggest miner at each step is the best choice  Alice can make in this strategy.

\newtheorem{thm1}{Theorem}
\begin{thm1}\label{thm1}
Consider two miners,  $m_1$ and $m_2$, with the computational powers  $P_{m1}$ and $P_{m2}$, respectively. Assuming that $P_{m1} > P_{m2}$, it is more profitable for the attacker to select $m_1$ as the target to add to chain $X$  than $m_2$.
\begin{IEEEproof}
Based on:
\setlength{\arraycolsep}{0.0em}
\begin{eqnarray}
&&P_{Y,F,i,\{m_2\}} > P_{Y,F,i,\{m_1\}}\nonumber\\
&&P_{X,S,i,\{m_2\}} < P_{X,S,i,\{m_1\}}\nonumber\\
&&or\nonumber\\
&&P_{X,S,i-1,\{m_2\}} < P_{X,S,i-1,\{m_1\}}\nonumber
\end{eqnarray}

and on \eqref{equ16}:
\setlength{\arraycolsep}{0.0em}
\begin{eqnarray}
R_{i,m_2}&{}\backsimeq{}&\frac{(P_{m_2} + \mu) \times P_{Y,F,i,\{m_2\}}}{\lambda \times (\frac{\mu+P_{m_2}}{\lambda-P_{m_2}})^{i+1}} \times F-F \nonumber\\
&&{=}\:\frac{(\lambda - P_{m_2}) \times P_{Y,F,i,\{m_2\}}}{\lambda \times P_{X,S,i-1,\{m_2\}}}  \times F-F \nonumber\\
&&{>}\:\frac{(\lambda - P_{m_1}) \times P_{Y,F,i,\{m_1\}}}{\lambda \times P_{X,S,i-1,\{m_1\}}}  \times F-F \nonumber\\
&&{=}\:\frac{(P_{m_1} + \mu) \times P_{Y,F,i,\{m_1\}}}{\lambda \times P_{X,S,i,\{m_1\}}} \times F-F \backsimeq R_{i,m_1}\nonumber
\end{eqnarray}
\begin{equation}
\Rightarrow R_{i,m_2} > R_{i,m_1}\nonumber
\end{equation}
in which $R_{i,m_1}$ is the required bribe for $m_1$ in state $i$.
\end{IEEEproof}
\end{thm1}

Theorem \ref{thm1} states that at each step, the biggest miner is the best choice to target. In {BFF}, Alice uses \textit{Algorithm 1} to calculate the  bribe amount at each state of the Markov chain, and miners use \textit{Algorithm 2} to choose the chain to mine on.
\begin{algorithm}[b]
 \caption{{Finding the required bribery for $m_j$ in state $i$}}
 \begin{algorithmic}[1]
  \FOR {each new Event}

  \STATE $j = C-i;$   ~ // $i$ is the Current State
  \IF{$i$ is not $V$ or $W$}
  \STATE Add $m_j$ ($j$th biggest miner) from $Y$ to $X;$
  \STATE $E$ = obtain $R_{i,m_j}$ for $m_j$ based on \eqref{equ16}
  \IF {(there is already bribery in the network)}
  \STATE Release ($E-existing \ bribe(s)$) as the bribe;
  \ELSE
  \STATE Release $E$ as the bribe;
  \ENDIF
  \ENDIF
  \ENDFOR
 \end{algorithmic}
 \end{algorithm}

\begin{algorithm}[t]
 \caption{{Determining the  profitable chain in state $i$}}
 \begin{algorithmic}[1]
  \FOR {each new Event}
  \IF{$i$ is not $V$ or $W$}   
  \STATE $R_i$ = current offered bribe;
  \IF {(equation \eqref{equ16} is valid)}
  \STATE $JoinX()$;
  \ELSE
  \STATE $JoinY()$;
  \ENDIF
  \ENDIF
  \ENDFOR
 \end{algorithmic}
 \end{algorithm}

In {BFF}, bribery in state $i$ is profitable for $m_{j;\forall j\leqslant C-i}$ (assuming that miners are indexed according to their computational power). Alice tries to attract new miner(s) to her chain as well as keeping the previous miners that were attracted in the previous states. Let us assume that $i$ is the current state. If the miners on   $Y$ find a block, the attack returns  to state $i+1$ and in this case, we assume that the miner who has joined  $X$ in  state $i$, will go back to $Y$. The attacker presumably offers bribe in state $i+1$ again, which will be the subject of Alice's decision in this new state. There are certain situations  the miner decides to stay on  $X$, but the consideration serves to establish an upper bound on the cost of the attack, since it underestimates the fraction of miners who might decide to mine on  $X$. Line 2 in $Algorithm\ 1$ states that there is no bribe for states before $i = C$.

\subsubsection{{CRB}  Strategy}
In CRB, Alice offers  a constant  bribe at each step of the attack, with commitment. For each event in  chain $X$, she releases the private key to a new account that has $K$ Bitcoins. At each state, miners check to see if the bribes from the current state to $V$ is profitable enough to join $X$. This strategy can persuade more miners to join if Alice chooses a proper value for $K$. By using \eqref{equ15}, miner $m$ can calculate the required bribe for each state, and find the  minimum value of $K$ for itself using \eqref{equ20}.

\begin{equation}
\label{equ20}
K_m = \frac{\sum_{i=CurrentState}^{V+1} N_{D_0i} \times R_{im}}{\sum_{i=CurrentState}^{V+1} N_{D_0i}} \geq K
\end{equation}
$K_m$ is the expected amount of constant bribe for $m$ who rationally uses \eqref{equ20} to decide. It is obvious that going forward along the chain towards state $V$ increases the success probability and makes $K$ profitable for other miners too.

\subsubsection{GVC Strategy}
In $BFF$, the uncertainty about attack's future increases its cost and  the average number of times the first few states are visited in the Markov chain.  As the amount of bribe in the initial attack states is high, frequent visits to them increase the cost significantly. GVC tries to give the miners the information needed to forecast the future through long-term calculation of the attack's profitability. GVC assures miners that there will be enough bribe in the coming states in order to keep mining on Alice's chain. This strategy attracts more miners at a lower cost.
In GVC, Alice can give off the chain guarantees (e.g. by smart contracts \cite{sayadjahanbin}) to assure miners that she will release the required bribe in the next states. This way, Alice is committed to release the private keys to the accounts holding the advertised  Bitcoin bribes, until the attack finishes.

In this strategy, when the attack starts, Alice announces the amount of bribe for each state. She releases her commitment to these bribes too. {CRB}  is a special case of GVC in which miners do not take other miner's rationality into account and the value of bribe is constant, i.e. $R_{i} = K; ~\forall i \leq C$.

Because of the commitments in GVC,  miners estimate the computational power distribution at each state by seeing which miner can be convinced by the bribe. Here, we assume that there is no bribery in the states beyond  $C$. The strategy tries to find a formula for $R_i$ to increase the attack success probability and to decrease its cost. Alice can  advertise a vector of bribes  rather than a formula. We denote this vector by $\bar R$. In GVC, miners calculate the probability of state transitions, and calculate the profitability of $X$ and $Y$. Then they choose a chain to mine on. The calculation of profitability is a two-step job that is shown in $Algorithm\ 3$.
Miners can pre-compute both steps of the algorithm. For the 2nd step, miners will have a vector of actions, i.e. $action_i$, that shows the action to be taken in state $i$. Each entry of the vector is  either $JoinX$ or $JoinY$.

\begin{algorithm}[h]
 \caption{Finding $NewMarkov$ and the profitable chain.}
 \begin{algorithmic}[1]
  \STATE 1ST STEP (Obtaining $NewMarkov$)
  \STATE $Miners$ = Sort the miners in descending order  ;
  \FOR {each state $i$}
  \STATE $List$ = 0;
  \FOR {each $m$ in $Miners$}
  \STATE // $m$ is the next biggest miner on $Y$;
  \IF{(The \eqref{equ16} is valid for $m$)}
  \STATE Add $m$ to $List$;
   \ELSE
  \STATE {Adjust the transition probabilities from   state $i$ (to $i-1$ \&\ $i+1$) based on the sum of  powers   in $List$;}  \STATE Break;
  \ENDIF
  \ENDFOR
  \ENDFOR
  \STATE Save the resultant probabilities in $NewMarkov$ and apply it to the primary Markov chain;
  \STATE // 2ND STEP (Obtaining the profitable chain)
  \FOR {each new Event}
  \STATE $i$ = Current State
  \IF{my power is not included in $i$ and $i$ is not $V$ or $W$}
  \STATE $R_i$ = Current proposed bribe(s);
  \IF {(The \eqref{equ15} is valid)}
  \STATE $JoinX()$;
  \ELSE
  \STATE $JoinY()$;
  \ENDIF
  \ENDIF
  \ENDFOR
 \end{algorithmic}
 \end{algorithm}

In the 2nd step, any miner can see if Alice's chain is profitable and join. To calculate the success probability as well as $AttackCost$, Alice needs the final Markov chain (or its transition matrix) which we denote it by $FinalMarkov$. In other words, she should determine which miners will choose her chain in each state beforehand. To this end, Alice runs the first step of  $Algorithm\ 3$ and determines $NewMarkov$ and then uses it to obtain $FinalMarkov$. To find $NewMarkov$ probability matrix, assume that $P_i$ is the {minimum} computational  power that could be brought to $X$ by $R_i$  in state $i$. Based on Theorem \ref{thm1}, all miners whose  powers are greater than or equal to $P_i$ will join $X$. The minimum value of $P_i$ can be obtained through the inequality of \eqref{equ16} by replacing $P_m$ with $P_i$ and solving it for this quantity.
Let $f(p)$ represent the computational power distribution of miners and $F(p)$ the cumulative distribution function. Running the first step of $Algorithm\ 3$ gives $\mu + 1 - F(P_i)$ as the  probability of state transition from $i$ to $i-1$. We {find} $NewMarkov$ as follows:
\begin{align}
\eta_i &= \mu + 1 - F(P_i)\\
\gamma_i &= F(P_i) - \mu\nonumber
\end{align}

Fig. \ref{fig:newmarkov} shows $NewMarkov$ after finishing the 1st step of $Algorithm\ 3$.
To obtain $FinalMarkov$ chain by running the 2nd step of $Algorithm\ 3$, we will do a similar job,  except  that \eqref{equ15} replaces \eqref{equ16} in the process.

If we write $NewMarkov$ in its Canonical Form, we will have $Q_{NewMarkov}$ (or $Q_{NM}$) and $R_{NM}$. For any miner $i,$ we define $\eta_j^i$ as follows:
\begin{align}
\eta_j^i &= \eta_j + x \times P_i ;  ~~~ x \in \{0,1\}\\
\gamma_j^i &= 1 - \eta_j^i\nonumber
\end{align}

where $x$ is $0$ when  miner $i$ whose computational power is $P_i$ is already added to $\eta_j$ in the first step of $Algorithm\ 3$, and $1$ otherwise. Now we define matrix $Q_{NM}^i$ and $R_{NM}^i$ and an initially zero $|Z| \times |S|$ matrix, namely $\mathfrak{Z}$ whose entry $\zeta_{ij}$ ($\forall i\ \in Z\ (Miners),\ \forall j\ \in S\ (States)$) is:

\begin{figure}[t]
        \centering
        \includegraphics[width=\columnwidth]{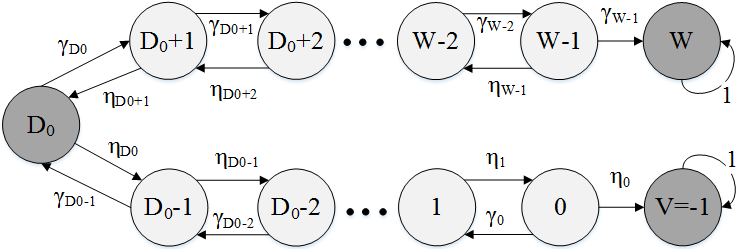}
        \vspace{-4ex}\caption{$NewMarkov$ chain, the result of the first step of $Algorithm\ 3$.}
        \label{fig:newmarkov}
\end{figure}

\begin{equation}
\label{equ21}
\zeta_{ij}= 
\begin{cases}
    1,~~& \text{if } (i\ is\ already\ added\ to\ X\ at\ state\ j)~ \text{or}\\
    &R_j\geqslant \frac{\{(I-Q_{NM})^{-1} \times R_{NM}\}_{j1} \times \eta_j^i}{\{(I-Q_{NM}^i)^{-1} \times R_{NM}^i\}_{j2} \times \gamma_j} \times F - F\\
    0,& \text{otherwise}
\end{cases}
\end{equation}
in which $\{...\}_{j1}$ means entry $[j][1]$ from the resultant matrix. We can use $\mathfrak{Z}$ and the miner's vector to calculate $FinalMarkov$. In \eqref{equ22}, $Tr[i]$ shows the transition probability from state $i$ to $i-1$ of $FinalMarkov$.
\begin{equation}
\label{equ22}
Tr = \mathfrak{Z} \times Z
\end{equation}

Now, we have $FinalMarkov$ that can be used to calculate the attack cost and success probability for GVC.

\subsection{The Future of Bribery Attack}
In the future, the bribery attack for cryptocurrencies like Bitcoin in which  reward goes down with  time, will be one of the most prominent attacks. Since the only factor needed to compare the profitability of two chains is their block rewards (because the transaction fee in the main chain can be {neutralized} by the transaction fee in the attacker's chain),  by  {halving} the block reward, the required bribe to persuade miners to mine on the attacker's chain will decrease too.

\subsection{Possible Countermeasures}
One solution to overcome this attack is limiting the amount of BTC that a user can transfer. The limitation is closely related to the amount of block reward. If the attacker's budget remains constant, decreasing the  reward increases the attack success chance. So, in some cryptocurrencies like Bitcoin,  mining algorithm should be equipped with a mechanism that adjusts the limit every 4 years (when the reward falls). However, this solution has a drawback,   since the attacker can transfer BTC with multiple addresses in the same block. 

The other solution for the drawback is limiting the  whole transferred BTC in each block. Like before, there should be a mechanism to adjust the limit in this solution. Such solution should take into account the situation in which there is no block reward and the only reward is the transaction fee.
In any case, at least a soft-fork is necessary.

\section{Numerical Analysis\label{sec:results}}
In our  analyses, we use the realistic mining power distribution  of the  Bitcoin pools available at \url{http://www.btc.com} (Table \ref{table:CompDist}). We assume Alice, i.e. the attacker, is  the largest of those pools ($\mu$ = 0.2123), and $P2$ is $m$ (the largest miner on $Y$).
We further assume that Alice has created at least one block on her chain during block confirmation on the main chain and also we assume that the states $i > C$ do not include any bribe. We analyze the following strategies:

\begin{itemize}
        \item \textit{Basic Strategy (BS)}: Only one rational miner (i.e. $m$) is targeted by Alice. This is similar to the Whale Attack.
        \item \textit{BFF}: Targeting the next biggest  miner on $Y$ in each state.
        \item \textit{CRB}: We analyze cases in which  bribery is done to attract $m$ as the biggest miner. Other miners may join $X$ in the states close to $V$, and our analysis will take those into account too. We will analyze two types of this strategy, CRB1 and CRB2. In CRB1, regardless which state the attack started, Alice calculates the value of $K$ for $m$ from state $C$. In CRB2 Alice starts calculating the bribe from $D_0$ and leaves the bribe for states $j;j>D_0$ zero.
        \item \textit{GVC}: To make it comparable with other strategies, we  assume that this strategy is used only to attract $m$ to Alice's chain. It is clear that Alice may decide to offer bribe in the final states based on $BaseMarkov$ to decrease the amount of bribe needed in the earlier states (to persuade $m$).
\end{itemize}

\begin{table}
\renewcommand{\arraystretch}{1}
\caption{Distribution of the mining power in Jul 2019 from https://btc.com.}
\label{table:CompDist}
\centering
\vspace{-1ex}\begin{tabular}{c|c|c|c|c|c}
\hline
\bfseries Abbr. & \bfseries Pool Name & \bfseries Power & \bfseries Abbr. & \bfseries Pool Name & \bfseries Power \\
\hline\hline
 & Whole Network & 1 & P8 & BTC.TOP & 0.0716\\
\hline
P1 & BTC.com & 0.2123 & P9 & BitFury & 0.0543 \\
\hline
P2 & AntPool & 0.1284 & P10 & Bitcoin.com & 0.0247\\
\hline
P3 & Poolin & 0.0988 & P11 & Huobi.pool & 0.0247\\
\hline
P4 & unknown & 0.0963 & P12 & BitClub & 0.0123\\
\hline
P5 & F2Pool & 0.0938 & P13 & WAYLCN & 0.0074\\
\hline
P6 & ViaBTC & 0.0864 & P14 & Bixin & 0.0049\\
\hline
P7 & SlushPool & 0.0815 & P15 & Eobot & 0.0025\\
\hline
\end{tabular}
\end{table}

\begin{figure}
        \centering
        \includegraphics[trim={0.6cm .2cm 0.9cm .4cm},width=\columnwidth]{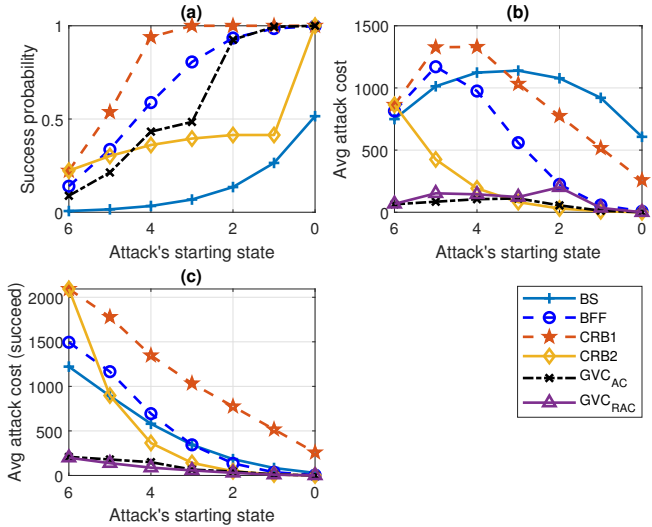}
        \vspace{-5ex}\caption{Success probability of different bribery attack strategies.}
        \label{fig:SuccProb}
\end{figure}

In all of the above strategies, we only focus on the results with minimum attack cost. Figure \ref{fig:SuccProb}(a) depicts the success probability of attack in different strategies when they try to attract $m$ to Alice's chain. In this figure, x-axis shows the starting state of the attack and y-axis shows the success probability. For example if the attack starts from  $i=4$, and Alice takes the $BFF$ strategy, then the success probability will be around \%60. The figure shows a low success probability for the $BS$ strategy compared to other strategies. All the results have been calculated from Alice's point of view. In this view, there is no bribing in the states $i > C$, and the transition probability remains constant. In $GVC$ and $CRB$, the miners' point of view is the same as Alice's, because the commitment discloses that there will be no bribe money in the states $i > C$. In other strategies, miners also counted themselves in for the states $i > C$ and during the calculation of the required bribe. They left Alice's chain when they saw that the chain was not profitable anymore. In these strategies, the success probability from the  miners' point of view is higher than those of Fig. \ref{fig:SuccProb}(a).

The other interesting result is the cost of attack in different strategies. Fig. \ref{fig:SuccProb}(b) shows the minimum of average cost regardless of the attack result, and, Fig. \ref{fig:SuccProb}(c) shows the minimum of average cost when the attack succeed. The figures show the minimum of average cost for each strategy when they try to attract $m$ to Alice's chain, and for different values of $i$ (starting state). For example, if $i = 4,$ the average cost of $CRB2$ will be about 192 BTC. These figures show a high average cost for the $BS$ strategy. Combined with Fig.~\ref{fig:SuccProb}(a), the results show the inefficiency of the $BS$ strategy. To calculate the attack cost (for BFF and BS, based on \eqref{equ18} and \eqref{equ19}), $R_i$ is obtained from the point of miner's view, and $N_{D_0i}$ from the point of attacker's view.

\begin{figure*}
       \centering
       \includegraphics[trim={0.5cm 0.7cm 1cm 0.7cm},width=\textwidth]{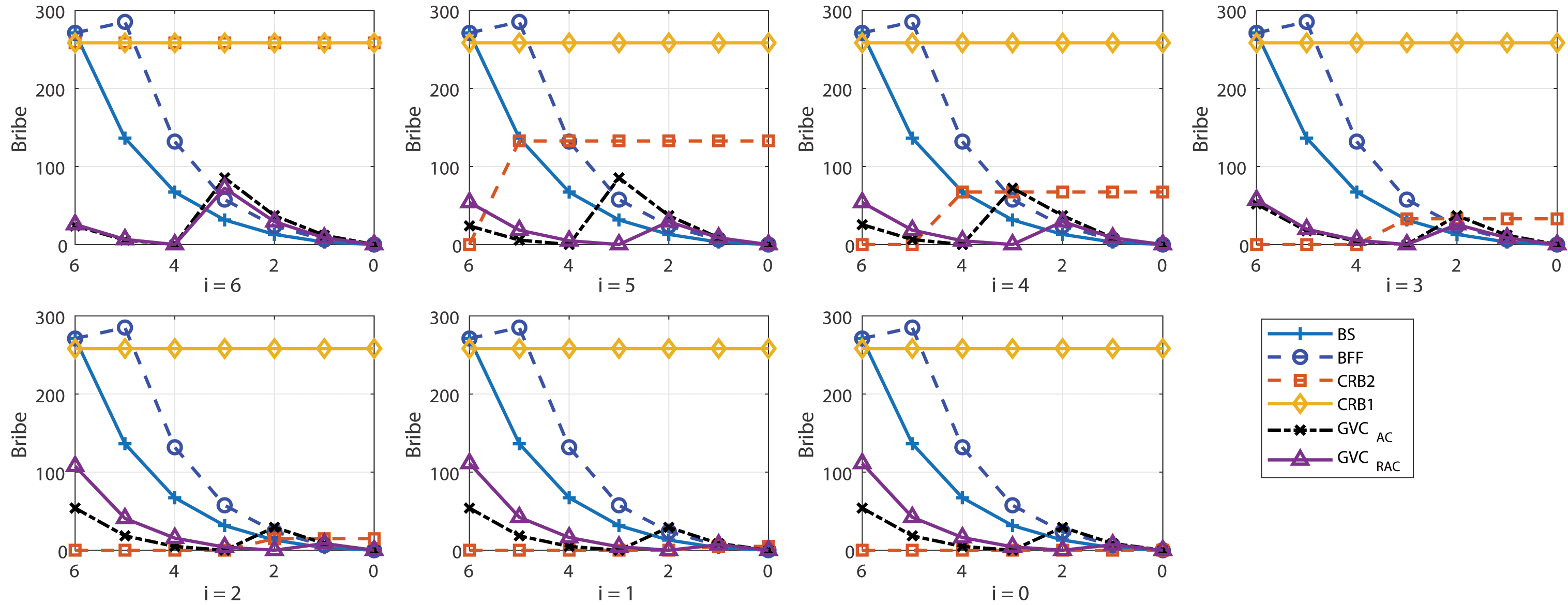}
       \vspace{-4.5ex}\caption{The amount of bribe at each state for different values of   $i$ as the starting state of the attack.}
       \label{fig:Bribe}
\end{figure*}


Fig. \ref{fig:Bribe} shows the bribe amount in each state for different values of $i$ (as the starting point of attack). It is clear that by increasing the bribe in each state for any strategy, the success probability of that strategy will go up. However, in the results, the calculation focuses only on the minimum cost of attack when Alice tries to attract $m$ as the biggest miner on the main chain. In the figure we have two types of $GVC$; $GVC_{AC}$ and $GVC_{RAC}$. The former focuses on the minimum average cost regardless of the attack result and the later focuses on the minimum average cost when the attack succeed. Based on the figure,  Alice should offer [25.51, 6.43, $\epsilon$, 72.25, 37.02, 8.6, $\epsilon$], where $\epsilon = One\ Satoshi$, as the bribe vector as well as commitment on the vector if she follows the $GVC_{AC}$ strategy. In this example, the average cost of attack and success probability will be 105 BTC and \%43.25, respectively.  In almost all cases, the $GVC$ strategy yields the lowest cost and offers reasonable success probabilities compared to other strategies. As shown in the figure, there is no change in bribe distribution for the $BS$, $BFF$ and $CRB1$ strategies with different values of $i$. Because these strategies only take specific miner(s) into account in each state thus letting other miners select the main chain as the profitable chain. For all $i$ values in Fig.~\ref{fig:Bribe}, the $GVC$ strategy tries to apply bribe based on $BaseMarkov$ in the final states. Such a distribution makes huge savings in the bribe spent in the earlier states. That is why in some states (e.g. in the middle states, like 2 or 3), the amount of bribe is higher than that of the earlier states. By comparing $GVC$ with $BS$ and $BFF$, the amount of bribes in the states $i=3$ and below, follow similar patterns, but increasing the   bribe in these states in $GVC$ and making commitment on them, decreases the required bribe for the states $i=4$ and above.

\begin{figure*}
       \centering
       \includegraphics[trim={2.5cm 0cm 2cm 0.4cm},width=\textwidth]{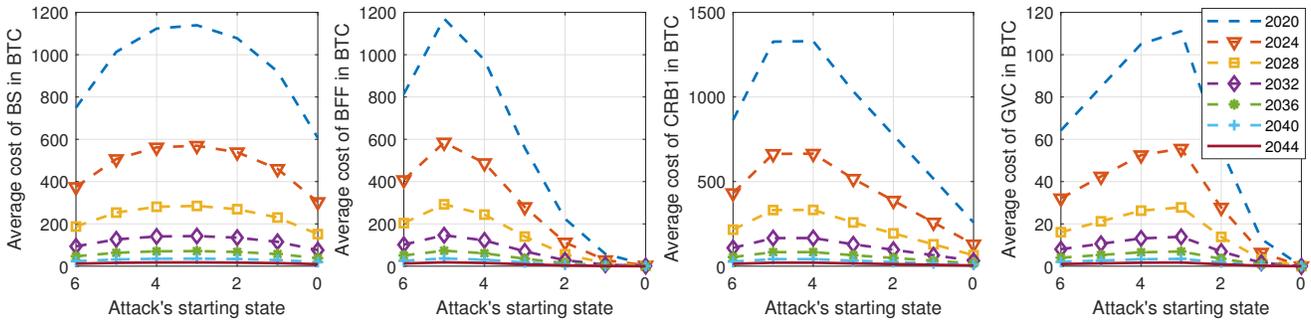}
              \vspace{-4.5ex}\caption{The attack cost in the future for different values of Bitcoin block reward}
       \label{fig:AttFuture}
\end{figure*}

In the previous figures, the amount of block reward is assumed to be 6.25 BTC.  Fig. \ref{fig:AttFuture} depicts the attack cost in the future for different values of Bitcoin block reward. As shown, the cost of attack falls down dramatically as the block reward is halved (rate 1/2). If the value of BTC remains almost constant (or its growth rate is equal to or less than 2), the attack cost  in Fig. \ref{fig:AttFuture} shows that  bribery attack will be a big challenge in the future of cryptocurrencies.

\section{Conclusion\label{sec:conclusion}}
In this paper, a novel analysis was done on the bribery attack which also generalized the previous analyses in this domain. Moreover, new strategies were presented to substantially reduce the cost of attack and increase its success probability.  It was shown that unlike the Whale Attack, an attacker can make bribery attack practical by designing and adopting right strategies.  The strategies   introduced, analyzed and simulated in this paper only serve as examples, yet they show how destructive this attack can be. The main contribution of the paper is paving the way by giving a rather general framework for the bribery attack that can accommodate any strategy.  The distribution of computational power in the cryptocurrency's network, influences the type of strategy adopted. It was shown that this attack will be a big challenge in the future of cryptocurrencies as the block rewards fade gradually. Two solutions were also proposed to mitigate the effect of this attack. The solutions  make soft-forks on Bitcoin thus it can remain backward-compatible.
 As a future work, we want to work on the solutions  proposed to bribery attack, and more specifically, on obtaining an effective limit on the amount of BTC in each transaction (1st solution) or in the whole block's transactions (2nd solution). These depend on some variables like block reward and level of confirmation in the target cryptocurrency.

\vspace{-1ex}

\bibliographystyle{IEEEtran}
\bibliography{referencesarxiv}
%



%

\vspace{+30ex}

\begin{IEEEbiography}[{\includegraphics[width=1in,height=1.25in,clip,keepaspectratio]{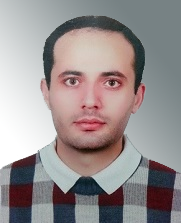}}]{Ghader Ebrahimpour} received the M.Sc. degree in Information Security from Amirkabir University of Technology (AUT) in 2015. He is also a member of Advanced Networking and Security research Laboratory (ANSLab).  Ghader has had several positions in industry before. He  is currently a Computer Science Ph.D. student at the University of Tehran (UT). His current research focuses on the analysis and design of cryptocurrency attacks on Blockchain-based systems.  \end{IEEEbiography}

\vspace{0ex}

\begin{IEEEbiography}[{\includegraphics[width=1in,height=1.25in,clip,keepaspectratio]{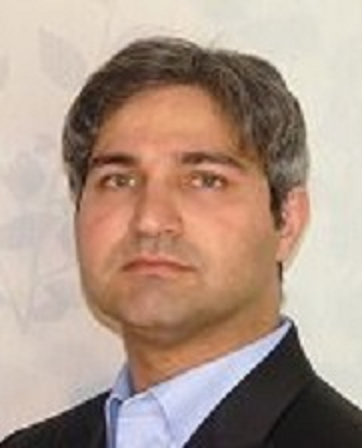}}]{Mohammad Sayad Haghighi} (IEEE SM'18) is the Head of IT Department at the University of Tehran, Iran. Prior to this, he was an Assistant Professor at Iran Telecom Research Center. He is also the director of Advanced Networking and Security research Laboratory (ANSLab). His research interests are wireless  networks  and cybersecurity. Dr. Sayad Haghighi has served as a PC member of many conferences such as IEEE WNS, IEEE SICK, IEEE HPCC, IEEE DASC, and IEEE LCN. He has won several national grants including some from Iran National Science Foundation (INSF).
\end{IEEEbiography}
\vfill







\end{document}